\documentclass[aps,prev,twocolumn,preprintnumbers,floatfix,nofootinbib]{revtex4-1}

\usepackage{graphicx}
\usepackage{amsmath, amsthm, amssymb}
\usepackage{amsfonts}
\usepackage{subfig}
\usepackage{xspace}
\usepackage[dvipsnames]{xcolor}
\usepackage{paralist}
\usepackage{multirow}
\usepackage{paralist}

\usepackage{graphicx}
\usepackage{bm}
\usepackage{times}
\usepackage{slashed}
\usepackage{color}
\usepackage{epsfig}\usepackage{dcolumn}
\usepackage{color}
\def\br{\begin{eqnarray}}
\def\er{\end{eqnarray}}
\def\be{\begin{equation}}
\def\ee{\end{equation}}

\begin{document}

\preprint{IIPDM-2019}

\title{Type I + II Seesaw in a Two Higgs Doublet Model}

\author{D. Cogollo$^1$}
\author{Ricardo D. Matheus$^2$}
\author{T\'{e}ssio B. de Melo$^{3,4}$}
\author{Farinaldo S. Queiroz$^{4}$}

\email{farinaldo.queiroz@iip.ufrn.br}

\affiliation{$^1$Departamento de F\'{\i}sica, Universidade Federal de Campina Grande, Campina Grande, PB, Brazil \\
$^2$Instituto de F\'isica Te\'{o}rica, Universidade Estadual Paulista, SP, Brazil\\
$^3$Departamento de F\'{\i}sica, Universidade Federal da Para\'\i ba, Caixa Postal 5008, 58051-970, Jo\~ao Pessoa, PB, Brazil \\
$^4$International Institute of Physics, Universidade Federal do Rio Grande do Norte, Campus Universit\'ario, Lagoa Nova, Natal-RN 59078-970, Brazil\\
}

\begin{abstract}
Two Higgs Doublet Models (2HDM) are popular extensions of the Standard Model for several reasons, but do not explain neutrino masses. In this work, we investigate how one can incorporate neutrino masses within the framework of the 2HDM-U(1), where U(1) is an abelian gauge symmetry used to nicely address the absence of flavor changing neutral currents in 2HDM. In particular, we explore realizations of the type I and type II seesaw since they are mechanisms that we dote on for being able to generate elegantly small active neutrino masses. We show that one can build several models featuring type I, type II and type I+II seesaw mechanism with different phenomenological implications.
\end{abstract} 
    
\maketitle

\section{Introduction}
\label{introduction}

The Standard Model (SM) has endured all precision tests in the past decades and offers the best description of the electroweak and strong interactions in nature \cite{Glashow:1961tr,Weinberg:1967tq}. The discovery of a scalar particle that resembles very much the SM Higgs has solidified it ever further\cite{Aad:2012tfae,Chatrchyan:2012xdj}. Neutrinos are massless in the SM though, which is in conflict with the observation of neutrino oscillations which require non-zero neutrino masses. Therefore the SM must be extended. The most trivial way to accommodate neutrino masses in the SM is via the introduction of right-handed neutrinos and then generate Dirac neutrino masses. The smallness of the neutrino masses would be explained by using suppressed Yukawa couplings. If in addition to the right-handed neutrinos we add a Majorana mass term for the right-handed neutrinos the type I seesaw mechanism arises. A Majorana mass term violates lepton number in two units, but lepton number is simply an accidental symmetry in the SM, so there is no fundamental reason that prohibits it \cite{Minkowski:1977sc,Mohapatra:1979ia}. The smallness of the active neutrino masses is explained by either invoking tiny Yukawa couplings or setting the Majorana mass at very high energy scales. Arguably the addition of a bare mass term means that the theory is not complete, and this bare mass term is expected to be related to a spontaneous symmetry breaking mechanism somehow \cite{Mohapatra:1980yp}. An orthogonal way to accommodate neutrino masses is by adding to the SM spectrum a scalar triplet, which features a neutral scalar with a very small vacuum expectation value that is responsible for generating neutrino masses at the eV scale. This setup leads to Majorana neutrinos and it is known as the type II seesaw \cite{Schechter:1980gr}. \\

That said, any attempt to explain neutrino masses via type I or type II seesaw mechanism requires extra scalars which may alter the SM predictions. One important parameter in this regard is the  $\rho$ parameter which connects the spontaneous symmetry breaking mechanism to the SM gauge bosons masses. It is constrained to be $\rho = 1.00039 \pm 0.00019$ \cite{Tanabashi:2018oca} and models that have an extended scalar sector might feature contributions to the W and Z masses that might bring $\rho$ away from the unit.\\ 

On one hand, Two-Higgs-Doublet Models (2HDM) are appealing because they naturally keep the $\rho$ parameter unchanged \cite{Lee:1973iz}. On the other hand, they fail to accommodate neutrino masses and for this reason, they should be extended if they stand at all as the theory beyond the SM.  Moreover, 2HDM in general feature flavor changing neutral interactions which are subject to stringent bounds and severely restrict the parameter space of such models \cite{Diaz:2002pa,Diaz:2002uk,Bai:2012ex,Altmannshofer:2012ar,Kim:2015zla}. \\

Some attempts have been made to improve the 2HDM by addressing dark matter \cite{LopezHonorez:2006gr,Gustafsson:2007pc,Dolle:2009fn,Honorez:2010re,LopezHonorez:2010tb,Bonilla:2014xba,Queiroz:2015utg,Arcadi:2018pfo,Bertuzzo:2018ftf}, neutrino masses \cite{Antusch:2001vn,Atwood:2005bf,Clarke:2015hta,Liu:2016mpf}, among others interesting observables. In this work we are interested in 2DHM  featuring an additional gauge symmetry. There are proposals in the literature involving non-abelian gauge groups which have interesting outcome \cite{Arhrib:2018sbz}, but in this manuscript we focus rather on 2HDM that are augmented by a $U(1)_X$ group motivated by the works done in \cite{Ko:2012hd,Heeck:2014qea,Huang:2015wts,Campos:2017dgc,DelleRose:2017xil}.\\

The initial motivation behind such models was that they could elegantly explain the absence of flavor changing interactions because the $U(1)$ gauge group could break down to a $Z_2$ symmetry that prevented both scalar doublets from contributing to fermion masses, and a richer collider phenomenology surfaced such as exotic Higgs decays involving a light $Z^\prime$, etc \cite{Ko:2013zsa,Ko:2014uka,Ko:2015fxa}. Later on, it was explicitly shown that the smallness of the neutrino masses could be simultaneously addressed in such models via the type I seesaw mechanism \cite{Ma:2000cc}. In the latter, many $U(1)_X$ models could be selected to explain neutrino masses. The possibility of explaining neutrino masses via a type II seesaw was explored in \cite{Ma:1998dx,Ma:2002nn,Grimus:2009mm,Camargo:2018uzw} were the it has shown that the viable $U(1)_X$ symmetries were severely restricted. Furthermore, knowing that dark matter constitutes a strong evidence for physics beyond the Standard Model~\cite{Acharya:2017ttl},  it was shown that a viable dark matter candidate via the well known $Z^\prime$ portal was feasible~\cite{Camargo:2019ukv}. The dark matter phenomenology was driven by the $Z^\prime$ interactions with dark matter and SM particles if dark matter acts as a vector-like fermion. Thus, qualitatively speaking, it does not matter whether we have a seesaw type I or type II at play. The models we propose in this work feature combinations of type I and II seesaw mechanisms, and since the dark matter phenomenology does not qualitatively depend on the seesaw mechanism one can easily accommodate a vector-like fermion as dark matter in the model proposed here. The phenomenology would be very similar and for this reason we will not repeat this exercise, and we will rather focus on the possible seesaw realizations.\\

In the type I seesaw mechanism three right-handed neutrinos are added to the 2HDM spectrum, and the mixing between the right-handed neutrinos and active neutrinos induces light masses to the active neutrinos after the diagonalization of the mass matrix. It is well-known that the addition of chiral fermions generates gauge anomalies that need to be cancelled out. In the proposal presented in \cite{Camargo:2019ukv} that was under control due to some gauge symmetries which played the same role as the usual B-L symmetry \cite{Rodejohann:2015lca} which requires the presence of three right-handed neutrinos to cancel the gauge anomalies. In the type II seesaw mechanism which invokes scalar a triplet, the gauge anomalies imply in severe restrictions to the SM quantum numbers limiting the possible $U(1)_X$ symmetries as aforementioned. In summary, we point out that there are several ways to accommodate neutrinos masses. The existence of right-handed neutrinos, a scalar triplet, a singlet scalar responsible for breaking the $U(1)_X$ allows different seesaw realizations, a fact that has not been explored in the past.  In our work, we review these aspects in a general setting and explore the connection to the absolute neutrino masses. In particular, we show that one can combine the type I and type II seesaw, and assess under which conditions one seesaw dominates over the other. Several type I+II seesaw studies have been performed in the past \cite{Ma:2002nn,Sui:2017qra,Bonilla:2017ekt,Biswas:2017dxt,Dev:2017ouk}, but in our work we discuss the type I, type II, and type I+II seesaw realizations embedded in the well motivated 2HDM-$U(1)$ model which has become an experimental benchmark at the LHC \cite{ATLAS:2018bvd,Abe:2018bpo,Farina:2018tju}, and investigate the implications for neutrino masses.\\

Our work is structured as follows: In section II we describe the 2HDM-$U(1)_X$ model, in section III we address the seesaw realizations, in section IV we discuss some phenomenological aspects before concluding in section V.

\begin{table*}[t]
\centering
\begin{tabular}{|c|c|c|c|c|c|}
\hline 
BM & Fields & Charge Assignment & Yukawa Lagrangian & Seesaw Type & Neutrino Nature \\ \hline \hline 
1& $ N _R , \Phi _s , \Delta $ & I & $y ^L \overline{L _L ^c} i \sigma ^2 \Delta L _L + y ^D \bar{L} _L \tilde{\Phi} _2 N _R + y ^R \overline{N _R ^c} \Phi _s N _R$ &  Type I + II &  Majorana   \\
2&  $ N _R , \Phi _s $ & I & $ y ^D \bar{L} _L \tilde{\Phi} _2 N _R + y ^R \overline{N _R ^c} \Phi _s N _R $ &  Type I &  Majorana   \\
3&  $ N _R , \Delta $ & I & $ y ^L \overline{L _L ^c} \Delta L _L + y ^D \bar{L} _L \tilde{\Phi} _2 N _R $ &  Type II + Dirac &  Majorana \\
4& $ N _R $ & I & $ y ^D \bar{L} _L \tilde{\Phi} _2 N _R $ &  Dirac   &  Dirac \\
5& \ $ \Phi _s , \Delta $ & II & $ y ^L \overline{L _L ^c} \Delta L _L $ & Type II &  Majorana \\
6& $ \Delta $ & II & $ y ^L \overline{L _L ^c} \Delta L _L $ & Type II & Majorana \\ \hline
\end{tabular}
\caption{Summary of the six general benchmark cases in this work where we investigate neutrino mass generation with and without the presence of right-handed neutrinos, a scalar triplet, and scalar singlet. Each scenario yields different scalar potentials and neutrino masses. See text for details.}
\label{table_seesaw_types_2hdm}
\end{table*}

\section{2HDM with $ U(1) _X $ symmetry}

Extending the SM via the inclusion of a second Higgs doublet amounts to the appearance of extra Yukawa interactions,
\begin{equation}
\begin{split}
\label{yukawa_2hdm_gen}
- \mathcal{L} _{ Y _{\text{2HDM}} } & = y _1 ^d \bar{Q} _L \Phi _1 d _R + y _1 ^u \bar{Q} _L \tilde{\Phi} _1 u _R + y _1 ^e \bar{L} _L \Phi _1 e _R \\
& + y _2 ^d \bar{Q} _L \Phi _2 d _R + y _2 ^u \bar{Q} _L \tilde{\Phi} _2 u _R + y _2 ^e \bar{L} _L \Phi _2 e _R + h.c. 
\end{split}
\end{equation}

The presence of these extra interaction terms can lead to flavor changing interactions mediated by extra neutral scalars at tree level. These flavor change neutral interactions (FCNI) should be suppressed in light of stringent bounds \cite{Cogollo:2013mga,Lindner:2016bgg,Batell:2017kty,Sirunyan:2017uae,Krnjaic:2019rsv}.  The standard way to prevent these processes is to impose a $Z _2$ symmetry, $\Phi _1 \rightarrow - \Phi _1$ and $\Phi _2 \rightarrow \Phi _2$, with suitable parity for the fermions, in order to eliminate the undesirable terms in equation (\ref{yukawa_2hdm_gen}). For instance, if all the fermions are even under $Z _2$, only the terms of the second line remain invariant,
\begin{equation}
\label{yukawa_2hdm_typeI}
- \mathcal{L} _{ Y _{\text{2HDM-I}} } = y _2 ^d \bar{Q} _L \Phi _2 d _R + y _2 ^u \bar{Q} _L \tilde{\Phi} _2 u _R + y _2 ^e \bar{L} _L \Phi _2 e _R + h.c. 
\end{equation} 
This Lagrangian characterizes the so-called type I 2HDM.  There are other types of Yukawa couplings which are free from FCNI and lead to different realizations known as type-II, flipped and lepton specific 2HDM \cite{Branco:2011iw}. In this work, we concentrate only with the type I 2HDM. Within this type I 2HDM, which has no neutrino masses, we can extend it using an abelian gauge symmetry and explain neutrino masses via a type I and type II seesaw as we describe further.\\

As said, an interesting alternative to using discrete symmetries to solve the flavor problem is by means of abelian gauge symmetries. The discrete symmetry was initially invoked to prevent one scalar doublet from contributing to fermion masses, and that can be elegantly done by imposing that one scalar doublet transforms differently from the other under a new gauge group. This symmetry can be used to solve the flavor problem, explain neutrinos masses and stabilize a potential dark matter candidate as aforementioned, its existence is much more appealing. Although, there are additional refinements that need to be made in order to have a consistent model.\\

The charges of the fields under the new gauge symmetry, $U(1) _X$, are constrained by the desired Yukawa interactions and triangle anomalies. That said, there are still several ways to accommodate neutrino masses, we will divide them into six benchmark scenarios.

\begin{itemize}
    \item {\bf BM 1}: Is the scenario where right-handed neutrinos, a scalar triplet and a scalar singlet are added to the 2HDM, inducing a type I +II seesaw mechanism;\\
    \item {\bf BM 2}: Concerns the setup where the 2HDM is augmented with only right-handed neutrinos and a scalar singlet, which leads to type I seesaw;\\
    \item {\bf BM 3}: In this case, in addition to three right-handed neutrinos a scalar triplet is invoked, yielding a type II seesaw;\\
    \item {\bf BM 4}: In this case only right-handed neutrinos are added to the 2HDM;
    \item {\bf BM 5}: Refers to the case where there are no right-handed neutrinos but singlet and triplet scalar fields are invoked;\\
    \item {\bf BM 6}: Is the setup where we simply add one scalar triplet.
\end{itemize}

We summarize these setups in Table~\ref{table_seesaw_types_2hdm} and
will describe each one in more detail below. 

If we simply augment the SM with an abelian gauge symmetry, all gauge anomalies can be cancelled without the need of extra chiral fermions. If we keep the fermion content of the SM, the cancellation of the $[U(1) _X]^3$ anomaly forces a relation between the charge of the right-handed up quarks ($u$), and the right-handed down quarks ($d$), namely $u = - 2d$. The charges of all the other fields can then be written in terms of one of them, say $d$, as follows:
\begin{equation}
\begin{split}
\label{expres_cargas_u}
q = - \frac{d}{2} & \text{\ \ , \ \ \ } l = \frac{3d}{2} \text{\ \ , \ \ \ } Q _{X _2} = - \frac{3d}{2} , \\
& u = - 2 d \text{\ \ , \ \ \ } e = 3 d ,
\end{split}
\end{equation}
where $l$ ($q$) is the $U(1) _X$ charge of the lepton (quark) doublet, $u$ ($e$) the charge of the right-handed up quarks (charged leptons) and $Q _{X _i}$ the charge of the scalar doublets. For later convenience, we will refer to it as {\it charge assignment II}. Note that the doublet $\Phi _1$ is neither coupled to fermions nor involved in gauge anomalies, so that its $U(1) _X$ charge remains unconstrained. At this point, the only requirement is that the charges of the scalar doublets under $U(1)_X$, namely $Q_{X1}$ and $Q_{X2}$, ought to be different,  i.e. $Q _{X1} \neq Q _{X2}$, in order to avoid FCNI. We will see later on that this is not always true. \\

Considering a different case, where three right-handed neutrinos are added to the SM spectrum, we notice that their inclusion can be parametrized by two $U(1)_X$ charges, chosen to be $u$ and $d$. The $[U(1) _X]^3$ anomaly which had previously enforced $u = - 2d$ in the scenario above, can now be cancelled out by simply making $n = - ( u + 2d )$, where $n$ is the $U(1) _X$ of the right-handed neutrinos. The other charges are given by,
\begin{equation}
\begin{split}
\label{expres_cargas_u_d}
q = \frac{1}{2} ( & u + d ) \text{\ \ , \ \ } l = - \frac{3}{2} ( u + d ) \text{\ \ , \ \ } Q _{X2} = \frac{1}{2} ( u - d ) \\
e & = - ( 2 u + d ) \text{\ \ , \ \ } n = - ( u + 2 d ).
\end{split}
\end{equation}
We will refer to this as {\it charge assignment I}, as can be seen in Table~\ref{table_seesaw_types_2hdm}.\\

In this class of models, the implementation of the seesaw mechanism for the generation of neutrino masses calls for the presence of extra scalar fields. With right-handed neutrinos charged under $U(1) _X$, a bare Majorana mass term $M _R \overline{N _R ^c} N _R$ is forbidden. Thus the type I seesaw mechanism cannot be realized. However, the inclusion of a scalar singlet $\Phi _s$, allows for the coupling,
\begin{equation}
\label{yukawa_neutrino_1}
- \mathcal{L} _\nu = y ^R \overline{N _R ^c} \Phi _s N _R + h.c. ,
\end{equation}
which, after spontaneous symmetry breaking of $U(1)_X$, generates a Majorana mass term. The quantum numbers of $\Phi _s$ under the symmetry group $SU(3) _c \times SU(2) _L \times U(1) _Y \times U(1) _X$ are $\Phi _s \sim (1, 1, 0, q _{X s}) $. Note that Eq. \eqref{yukawa_neutrino_1} fixes the $U(1) _X$ charge of $\Phi _s$ as $q _{X s} = 2u + 4d$.\\

If right-handed neutrinos are not included, neutrino masses can still be generated provided that we add to the model a scalar triplet $\Delta \sim (1, 3, 2, q _{X t})$, so that the Yukawa coupling,
\begin{equation}
\label{yukawa_neutrino_2}
- \mathcal{L} _\nu = y ^L \overline{L _L ^c} i \sigma ^2 \Delta L _L + h.c. ,
\end{equation}
generates a Majorana mass term for the neutrinos after $\Delta$ acquires a vacuum expectation value ($vev$), which is the key signature of the type II seesaw mechanism. The term in eq.\eqref{yukawa_neutrino_2} is only present if the $U(1) _X$ charge of $\Delta$ is $q _{X t} = - 3d$. \\

In summary, one can generate neutrino masses through type I and/or type II seesaw mechanisms, and exploit this fact considering all possible realizations in the 2HDM-$U(1)_X$ framework.\\

\section{Seesaw realizations in the 2HDM-U(1)}

In the type I seesaw mechanism, heavy right-handed neutrinos lead to small neutrino masses, whereas in the type  II seesaw the small scalar triplet $vev$ justifies the small neutrino masses \cite{Nomura:2017emk,Pires:2018kaj,Vien:2018otl,Ouazghour:2018mld,Antusch:2018svb,Du:2018eaw,Parida:2018apw,Dev:2018kpa,Rodrigues:2018jpv,Dev:2018sel,Li:2018abw,Agrawal:2018pci,Ohlsson:2019sja}. Notice that these mechanisms are completely different from one another but at the end have the same goal. There are several ways to incorporate them in the 2HDM-$U(1)_X$ model, with or without right-handed neutrinos, a scalar triplet, and a scalar singlet field. Although, they lead only to two different charge assignments for the SM fields, we highlight that each scenario corresponds to a different model. In this work we extend previous studies by proposing new models where these seesaw realizations successfully happen. We will describe them below.  \\

\subsection{Type I + II seesaw mechanism ({\bf BM 1})}
\label{sub_sec_1}

It is possible to merge the Type I and Type II seesaw mechanisms by including both the scalar singlet and triplet. As right-handed neutrinos are also included, the charges follow the charge assignment I, under which the charge of the triplet is $q _{Xt} = 3 (u + d) $. In this general case, the Yukawa Lagrangian relevant for neutrino masses is given by,
\begin{equation}
\label{yukawa_lag_neutrino_masses}
- \mathcal{L} _\nu = y ^L \overline{L _L ^c} i \sigma ^2 \Delta L _L + y ^D \bar{L} _L \tilde{\Phi} _2 N _R + y ^R \overline{N _R ^c} \Phi _s N _R + h.c. 
\end{equation}
As the scalars develop their respective $vev$, the neutrinos acquire masses according to  
\begin{equation}
\label{neutrino_mass_lagrangian}
- \mathcal{L} _\nu = \frac{1}{2} \overline{\nu _L ^c} M _L \nu _L + \bar{\nu} _L M _D N _R + \frac{1}{2} \overline{N _R ^c} M _R N _R + h.c. ,
\end{equation}
with,
\begin{equation}
\label{mass_matrices}
\frac{1}{2} M _L = \frac{y ^L v _t}{\sqrt{2}} \text{\ \ \ , \ \ \ } M _D = \frac{y ^D v _2}{\sqrt{2}} \text{\ \ \ , \ \ \ } \frac{1}{2} M _R = \frac{y ^R v _s}{\sqrt{2}} ,
\end{equation}
where $v_t$, $v_2$ and $v_s$ are respectively the $vev$ of
$\Delta$, $\Phi_2$ and $\Phi_s$.
Although we have suppressed the flavor indices, it is to be understood that $M _L$, $M _R$, $M _D$ and the corresponding Yukawa couplings are $3 \times 3$ matrices in flavor space.\\

We can arrange the left-handed active neutrinos and right-handed ones in a left-handed neutrino field as,
\begin{equation}
N _L = \begin{pmatrix} \nu _L \\ N _R ^c \end{pmatrix} ,
\end{equation}and rewrite Eq.\eqref{neutrino_mass_lagrangian} in a matrix form,
\begin{equation}
\label{neutrino_mass_lagrangian_matrix_form}
- \mathcal{L} _\nu = \frac{1}{2} \overline{N _L ^c} M _\nu N _L + h. c. ,
\end{equation}
with the mass matrix 
\begin{equation}
M _\nu = \begin{pmatrix} M _L & M _D ^T \\ M _D & M _R \end{pmatrix} ,
\end{equation}
whose eigenvalues give the physical neutrino masses.\\

As we are interested in estimating the order of magnitude of the physical neutrino masses, 
we will use the simplifying assumption that the matrices $M _L$, $M _R$ and $M _D$ are diagonal, i.e., $M _L = \text{diag} ( m _L , m _L , m _L ) $, $M _R = \text{diag} ( m _R , m _R , m _R ) $ and  $M _D= \text{diag} ( m _D , m _D , m _D ) $, where these masses are real and positive. Consequently, $M _\nu$ reads,
\begin{equation}
\label{nu_mass_matrix_simplyfied}
M _\nu = \begin{pmatrix} m _L & 0 & 0 & m _D & 0 & 0 \\ 0 & m _L & 0 & 0 & m _D & 0 \\ 0 & 0 & m _L & 0 & 0 & m _D \\ m _D & 0 & 0 & m _R & 0 & 0 \\ 0 & m _D & 0 & 0 & m _R & 0 \\ 0 & 0 & m _D & 0 & 0 & m _R \end{pmatrix} ,
\end{equation}
and its eigenvalues are degenerate and given by, 
\begin{equation}
m  , M = \frac{1}{2} \left[ m _L + m _R \mp \sqrt{ 4 m _D ^2 + ( m _L - m _R ) ^2 } \right] ,
\end{equation}
where the minus (plus) sign corresponds to 
neutrino masses $m$ ($M$). It should be clear that there are six eigenvalues actually, three of them equal to $m$ and the others equal to $M$. This degeneracy is a result of our simplifying assumption on $M _\nu$. Obviously, this scenario of mass degenerate neutrinos does not reproduce the neutrino oscillation data, but that can be easily achieved by letting $M_L$ and $M_D$ not be diagonal as shown in \cite{Ferreira:2019qpf}. 

Depending on the relative sizes of $m _D$, $m _R$ and $m _L$, there are several distinct scenarios for the neutrino masses. 
In the Table \ref{mass_limits_2hdm} approximate expressions for them are summarized, and the explicit derivation is shown in the {\it appendix}. We see that the first four cases in Table \ref{mass_limits_2hdm} commonly feature $ m _R \gg m _D $, i.e. the 
neutrino masses are essentially given by $m = m _L$ ($M = m _R$), so that to obtain active neutrino masses of order $\sim 0.1$ eV, $m _L$ is forced to be very small, $m _L \lesssim 0.1$  eV.\\

In the next two rows which assume $m_D\gg m_R$, all the neutrinos are practically mass degenerate, with masses set by $m _D$,  and are known as pseudo-Dirac neutrinos \cite{Geiser:1998mr,Chang:1999pb,Beacom:2003eu}. This scenario however is not realistic because, on one hand, CMB data constrains the sum of active neutrino masses \cite{Ade:2015xua},
\begin{equation}
\sum _i m _i \lesssim 0.1 \text{ eV},
\end{equation}and on the other hand, stable right-neutrinos behave like dark matter, and successful structure formation impose, \cite{Safarzadeh:2018hhg,Martins:2018ozw,Villanueva-Domingo:2017lae,Lopez-Honorez:2017csg,Irsic:2017ixq},
\begin{equation}
M \gtrsim 1 \text{ keV} , 
\end{equation}
ruling out this kind of pseudo-Dirac neutrinos. Nevertheless, if the right-handed neutrinos are unstable particles, then the bounds can be avoided and, in principle, would be possible to have $M$ as low as $0.1$ eV. Specifically in our model, this possibility could only be realized through the decay channel enabled by the Yukawa interaction,
\begin{equation*}
- \mathcal{L} _\nu \supset y ^D \bar{L} _L \tilde{\Phi} _2 N _R = \frac{y ^D v _2}{\sqrt{2}} \bar{\nu} _L N _R + \frac{y ^D}{\sqrt{2}} \bar{\nu} _L \rho _2 N _R ,
\end{equation*}
where $\rho _2$ is the CP-even scalar of the $\Phi _2$ doublet. However, with such a small mass of $N _R$ this decay becomes kinematically forbidden, what makes right-handed neutrinos stable in our model, conclusively excluding this scenario.\\

In the last row, $m _D$ and $m _R$ being of the same order of magnitude imply that $m$ and $M$ are also of the same order or magnitude, but with $m$ being slightly smaller than $M$, unless $m _D$ and $m _R$ are finely tuned. Therefore, this scenario is similar to the previous pseudo-Dirac case, in other words, ruled out.\\

\begin{table*}[t]
\centering
\begin{tabular}{c|c|c|c}
\hline
Limit & $ m $ & $ M $ & Neutrino Nature \\ \hline \hline 
$ m _R \gg m _D \gg m _L$ & $m _L - \frac{m _D ^2}{2 m _R}$ & $m _R + \frac{m _D ^2}{2 m _R}$ & Majorana \\
$ m _R \gg m _L \gg m _D $  & $m _L - \frac{m _L ^2}{4 m _R}$ & $m _R + \frac{m _L ^2}{4 m _R}$ & Majorana \\
$ m _R \gg m _D , m _L$ and $ m _D \sim m _L $ & $m _L - \frac{m _D ^2}{2 m _R} - \frac{m _L ^2}{4 m _R}$ & $m _R + \frac{m _D ^2}{2 m _R} + \frac{m _L ^2}{4 m _R}$ & Majorana \\
$ m _D \ll m _R , m _L $ and $ m _R \sim m _L $ &  $m _L - \frac{m _D ^2}{( m _R - m _L )}$ & $m _R + \frac{m _D ^2}{( m _R - m _L )}$ & Majorana \\
$ m _D \gg m _R \gg m _L$ &  $- m _D + \frac{1}{2} m _R - \frac{m _R ^2}{8 m _D}$ & $m _D + \frac{1}{2} m _R + \frac{m _R ^2}{8 m _D}$ & Pseudo-Dirac \\
$ m _D \gg m _R , m _L$ and $ m _R \sim m _L $ &  $- m _D + \frac{m _L + m _R}{2} - \frac{( m _L - m _R ) ^2}{8 m _D}$ & $m _D + \frac{m _L + m _R}{2} + \frac{( m _L - m _R ) ^2}{8 m _D}$ & Pseudo-Dirac \\
$ m _L \ll m _R , m _D $ and $ m _R \sim m _D $ & $ \frac{1}{2} \left[ m _R - \sqrt{ 4 m _D ^2 + m _R ^2 } \right] $ & $ \frac{1}{2} \left[ m _R + \sqrt{ 4 m _D ^2 + m _R ^2 } \right] $ & Pseudo-Dirac  \\ \hline
\end{tabular}
\caption{{\footnotesize Physical neutrino masses in different limits of the type I + II seesaw mechanism in 2HDM (bechmark scenario {\bf BM 1}).}}
\label{mass_limits_2hdm}
\end{table*}

Each one of the cases discussed previously will lead to a different scalar potential that we describe further. We remind the reader that we are focused on the 2HDM-$U(1)_X$ models where the doublet $\Phi _1$ does not couples to the SM fermions. Therefore, there is freedom to choose different $U(1) _X$ charges, consequently leading to various scalar potentials. In general, the scalar potential can be written as,
\begin{equation}
\begin{split}
V ( \Phi _1 , & \Phi _2 , \Phi _s , \Delta ) = V _H + V _{NH} ,
\end{split}
\end{equation}
where $V _H$ stands for the part of the potential that contains Hermitian terms,
\begin{widetext}
\begin{equation}
\begin{split}
\label{potential_doblets_triplet_singlet_u1_x_with_Nr}
V _H & = m _1 ^2 \Phi _1 ^\dagger \Phi _1 + m _2 ^2 \Phi _2 ^\dagger \Phi _2 + m _s ^2 \Phi _s ^\dagger \Phi _s + m _t ^2 \text{Tr} ( \Delta ^\dagger \Delta ) + \lambda _1 ( \Phi _1 ^\dagger \Phi _1 ) ^2 + \lambda _2 ( \Phi _2 ^\dagger \Phi _2 ) ^2 + \lambda _s ( \Phi _s ^\dagger \Phi _s ) ^2 + \lambda _t [ \text{Tr} ( \Delta ^\dagger \Delta ) ] ^2 \\
& + \lambda _{tt} \text{Tr} ( \Delta ^\dagger \Delta ) ^2 + \lambda _3 ( \Phi _1 ^\dagger \Phi _1 ) ( \Phi _2 ^\dagger \Phi _2 ) + \lambda _4 ( \Phi _1 ^\dagger \Phi _2 ) ( \Phi _2 ^\dagger \Phi _1 ) + \lambda _{s1} ( \Phi _1 ^\dagger \Phi _1 ) ( \Phi _s ^\dagger \Phi _s ) + \lambda _{s2} ( \Phi _2 ^\dagger \Phi _2 ) ( \Phi _s ^\dagger \Phi _s ) \\
& + \lambda _{t1} ( \Phi _1 ^\dagger \Phi _1 ) \text{Tr} ( \Delta ^\dagger \Delta ) + \lambda _{t2} ( \Phi _2 ^\dagger \Phi _2 ) \text{Tr} ( \Delta ^\dagger \Delta ) + \lambda _{tt1} \Phi _1 ^\dagger \Delta \Delta ^\dagger \Phi _1 + \lambda _{tt2} \Phi _2 ^\dagger \Delta \Delta ^\dagger \Phi _2 + \lambda _{st} ( \Phi _s ^\dagger \Phi _s ) \text{Tr} ( \Delta ^\dagger \Delta )
\end{split}
\end{equation}
\end{widetext}
and $V _{NH}$ corresponds to the remaining non-Hermitian ones.

There are three possibilities, depending on the charge of $\Phi _1$, $Q _{X1}$. These three possibilities rise after considering the Yukawa lagrangians which should remain intact. They read,

{\bf (i)} for $Q _{X1} = \frac{1}{2} ( 5u + 7d ) $ we get,
\begin{equation}
\begin{split}
V _{NH} & = \mu _s ( \Phi _1 ^\dagger \Phi _2 \Phi _s + h. c. ) + \mu _t ( \Phi _1 ^T i \sigma ^2 \Delta ^\dagger \Phi _2 + h.c. ) \\
& + \kappa _1 ' ( \Phi _1 ^T i \sigma ^2 \Delta ^\dagger \Phi _1 \Phi _s ^\dagger + h.c. ) \\
& + \kappa _2 ( \Phi _2 ^T i \sigma ^2 \Delta ^\dagger \Phi _2 \Phi _s + h.c. ) ;
\end{split}
\end{equation}

{\bf (ii)} for $Q _{X1} = \frac{3}{2} ( u + d ) $ we find,
\begin{equation}
V _{NH} = \mu _{t1} ( \Phi _1 ^T i \sigma ^2 \Delta ^\dagger \Phi _1 + h.c. ) + \kappa _2 ( \Phi _2 ^T i \sigma ^2 \Delta ^\dagger \Phi _2 \Phi _s + h.c. ) ;
\end{equation}

{\bf (iii)} and for $Q _{X1} = \frac{3}{2} ( 3u + 5d ) $:
\begin{equation}
V _{NH} = \kappa ' ( \Phi _1 ^T i \sigma ^2 \Delta ^\dagger \Phi _2 \Phi _s ^\dagger + h.c. ) + \kappa _2 ( \Phi _2 ^T i \sigma ^2 \Delta ^\dagger \Phi _2 \Phi _s + h.c. ) .
\end{equation}

Notice that indeed there are three different distinct non-Hermitian scalar potentials which can be further modified depending on the presence or not of the scalar triplet and singlet field. We will consider these cases below.\\

\subsection{Scalar singlet absent ({\bf BM 3})}
\label{sub_sec_2}

In this section we shall consider the case in which the scalar sector is composed only by the doublets $\Phi _i$ and the triplet $\Delta$. Without the scalar singlet $\Phi _s$, the last term in equation (\ref{neutrino_mass_lagrangian}) is absent, which amounts to a vanishing $M _R$, so that,
\begin{equation}
M _\nu = \begin{pmatrix} M _L & M _D ^T \\ M _D & 0 \end{pmatrix} .
\end{equation}

The eigenvalues of this matrix are,
\begin{equation}
m = \frac{1}{2} \left[ \sqrt{ 4 m _D ^2 + m _L ^2 } - m _L \right] ,
\end{equation}
and,
\begin{equation}
M = \frac{1}{2} \left[ m _L + \sqrt{ 4 m _D ^2 + m _L ^2 } \right] .
\end{equation}

In this setup there are {\it three} variants, summarized in table~\ref{mass_limits_2hdm2}. The first possibility is $ m _D \gg m _L $. In this limit we get,
\begin{equation*}
m , M \simeq m _D \mp \frac{1}{2} m _L + \frac{m _L ^2}{8 m _D} ,
\end{equation*}which approximately means that,
\begin{equation}
m , M \simeq m _D .
\end{equation}

Thus the neutrinos are pseudo-Dirac neutrinos, and as we discussed previously, this scenario is excluded. \\

The {\it second} possibility happens when $m _L \sim m _D $. If $m _L$ and $m _D$ are of the same order of magnitude the same happens for $m$ and $M$, but with $m$ being slightly smaller than $M$, unless again we invoke some fine tuning. \\ 

The {\it third} case occurs for $m _L \gg m _D$, which leads to,
\begin{equation}
\label{left_handed_nu_mass_no_singlet_ml__md}
m \simeq \frac{m _D ^2}{m _L},
\end{equation}
and,
\begin{equation}
M \simeq m _L .
\end{equation}
From eq.\eqref{left_handed_nu_mass_no_singlet_ml__md} we see that $m$ can be very small for sufficiently large $m _L$. However, we must take into account the constraints coming from the $\rho$ parameter, which preclude the $vev$ of the scalar triplet take on high values, thus limiting the maximum value of $m _L$. We can expect $m _L \lesssim 1$ GeV as a reasonable upper limit. Therefore, the only way to achieve $m _L \gg m _D$ is to make the Yukawa couplings $y ^D$ very small. For example, assuming $v _2 \sim 100$ GeV and $m _L \sim 100$ MeV, we need $y ^D \sim 10 ^{-8}$ to obtain $m \sim 0.1$ eV. Here, the right-handed neutrinos would have masses of $M \sim 100$ MeV. Right-handed neutrinos with masses around $100$~MeV are fully consistent with structure formation bounds if they are potential dark matter candidates \cite{Hooper:2003sh}. \\

\begin{table*}[t]
\begin{tabular}{c|c|c|c}
\hline
Limit & $ m $ & $ M $ & Neutrino Nature \\ \hline \hline 
$ m _D \gg m _L$ & $ m _D $ & $ m _D $ & pseudo-Dirac \\
$ m _D \sim m _L $  & $ \frac{1}{2} [ \sqrt{ 4 m _D ^2 + m _L ^2 } - m _L ] $ & $ \frac{1}{2} [ m _L + \sqrt{ 4 m _D ^2 + m _L ^2 } ] $ & Majorana \\
$ m _D \ll m _L $  & $ m _D ^2 / m _L $ & $ m _L $ & Majorana \\ \hline
\end{tabular}
\caption{Physical neutrino masses in different limits of type II seesaw mechanism of benchmark scenario {\bf BM 3} in 2HDM.}
\label{mass_limits_2hdm2}
\end{table*}

In this case, the Hermitian part of the potential is the same as the one in Eq. \eqref{potential_doblets_triplet_singlet_u1_x_with_Nr}, omitting the terms which contain the singlet:
\begin{widetext}
\begin{equation}
\begin{split}
\label{potential_doblets_triplet_u1_x_with_Nr}
V _H & = m _1 ^2 \Phi _1 ^\dagger \Phi _1 + m _2 ^2 \Phi _2 ^\dagger \Phi _2 + m _t ^2 \text{Tr} ( \Delta ^\dagger \Delta ) + \lambda _1 ( \Phi _1 ^\dagger \Phi _1 ) ^2 + \lambda _2 ( \Phi _2 ^\dagger \Phi _2 ) ^2 \\
& + \lambda _t [ \text{Tr} ( \Delta ^\dagger \Delta ) ] ^2 + \lambda _{tt} \text{Tr} ( \Delta ^\dagger \Delta ) ^2 + \lambda _3 ( \Phi _1 ^\dagger \Phi _1 ) ( \Phi _2 ^\dagger \Phi _2 ) + \lambda _4 ( \Phi _1 ^\dagger \Phi _2 ) ( \Phi _2 ^\dagger \Phi _1 ) \\
& + \lambda _{t1} ( \Phi _1 ^\dagger \Phi _1 ) \text{Tr} ( \Delta ^\dagger \Delta ) + \lambda _{t2} ( \Phi _2 ^\dagger \Phi _2 ) \text{Tr} ( \Delta ^\dagger \Delta ) + \lambda _{tt1} \Phi _1 ^\dagger \Delta \Delta ^\dagger \Phi _1 + \lambda _{tt2} \Phi _2 ^\dagger \Delta \Delta ^\dagger \Phi _2 .
\end{split}
\end{equation}
\end{widetext}

Regarding the non-Hermitian part of the potential there are some possibilities depending on the charge of $ Q _{X1} $. Two straightforward possibilities are $Q _{X1} = \frac{1}{2} ( 5u + 7d )$ that yields,
\begin{equation}
V _{NH} = \mu _t ( \Phi _1 ^T i \sigma ^2 \Delta ^\dagger \Phi _2 + h.c. ) , 
\end{equation}and $Q _{X1} = \frac{3}{2} ( u + d ) $ whic leads to,
\begin{equation}
V _{NH} = \mu _{t1} ( \Phi _1 ^T i \sigma ^2 \Delta ^\dagger \Phi _1 + h.c. ) .
\end{equation}
There is also a less obvious third option in which $Q _{X1}$ remains free and $u$ and $d$ are not independent anymore, but satisfy $u = - 2d$:
\begin{equation}
V _{NH} = \mu _{t2} ( \Phi _2 ^T i \sigma ^2 \Delta ^\dagger \Phi _2 + h.c. ) .
\end{equation}

The condition $u = - 2d$ requires the scalar singlet to be neutral under the $U(1) _X$ symmetry, and thus it cannot break this symmetry spontaneously. However, as we are not including the singlet here, we do not have to worry about this. Note also that the condition $u = - 2d$ forces the right-handed neutrinos to have zero $U(1)_X$ charges. Consequently, the bare mass term $M _R \overline{N _R ^c} N _R$ is now allowed going back to the case where a right-handed mass term is present. Albeit, the situation is fundamentally different because the entries of the matrix $M _R$ are free parameters.\\

\subsection{Scalar triplet absent - Type I seesaw ({\bf BM 2})}
\label{sub_sec_3}

Without the presence of the scalar triplet, the first term in equation (\ref{yukawa_lag_neutrino_masses}) is absent, so that, 
\begin{equation}
- \mathcal{L} _{ Y _{N _R} } = y _2 ^D \bar{L} _L \tilde{\Phi} _2 N _R + y ^M \overline{N _R ^c} \Phi _s N _R + h.c. 
\end{equation}
After spontaneous symmetry breaking, the Dirac and Majorana mass terms leads to the following mass matrix,
\begin{equation}
M _\nu = \begin{pmatrix} 0 & M _D ^T \\ M _D & M_R \end{pmatrix} 
\end{equation}
The physical neutrino masses are,
\begin{equation}
m , M = \frac{1}{2} \left[ m _R \pm \sqrt{4 m _D ^2 + m _R ^2} \right] .
\end{equation}

In the limit $m _R \gg m _D$, the type I seesaw mechanism is realized, so that,
\begin{equation}
m \simeq \frac{m _D ^2}{m _R} ,
\end{equation}
\begin{equation}
M \simeq m _R .
\end{equation}

In this scenario the scalar potential is uniquely defined with,
\begin{equation}
\begin{split}
\label{potential_doblets_singlet_u1_x_with_Nr}
V & = m _1 ^2 \Phi _1 ^\dagger \Phi _1 + m _2 ^2 \Phi _2 ^\dagger \Phi _2 + m _s ^2 \Phi _s ^\dagger \Phi _s + \lambda _1 ( \Phi _1 ^\dagger \Phi _1 ) ^2 \\
& + \lambda _2 ( \Phi _2 ^\dagger \Phi _2 ) ^2 + \lambda _s ( \Phi _s ^\dagger \Phi _s ) ^2 + \lambda _3 ( \Phi _1 ^\dagger \Phi _1 ) ( \Phi _2 ^\dagger \Phi _2 ) \\
& + \lambda _4 ( \Phi _1 ^\dagger \Phi _2 ) ( \Phi _2 ^\dagger \Phi _1 ) + \lambda _{s1} ( \Phi _s ^\dagger \Phi _s ) ( \Phi _1 ^\dagger \Phi _1 ) \\
& + \lambda _{s2} ( \Phi _s ^\dagger \Phi _s ) ( \Phi _2 ^\dagger \Phi _2 ) + \mu _s ( \Phi _1 ^\dagger \Phi _2 \Phi _s + h.c. ) ,
\end{split}
\end{equation}where $Q _{X1} = \frac{1}{2} ( 5u + 7d )$.

\subsection{Scalar singlet and triplet absent - Dirac neutrinos ({\bf BM 4})}
\label{sub_sec_4}

In the 2HDM without extra scalars, the Yukawa Lagrangian reduces to,  
\begin{equation}
- \mathcal{L} _{ Y _{N _R} } = y _2 ^D \bar{L} _L \tilde{\Phi} _2 N _R + h.c. .
\end{equation}
In this case, the neutrinos are Dirac particles and acquire mass similarly to the other SM fermions,
\begin{equation}
m = \frac{y _2 ^D v _2}{\sqrt{2}} .
\end{equation}

In this case, the smallness of neutrino masses requires small Yukawa couplings, as it happens when the SM is simply augmented by right-handed neutrinos. The scalar potential is given by,
\begin{equation}
\begin{split}
\label{potential_doblets_u1_x_with_Nr}
V ( \Phi _1 , \Phi _2 ) & = m _1 ^2 \Phi _1 ^\dagger \Phi _1 + m _2 ^2 \Phi _2 ^\dagger \Phi _2 + \lambda _1 ( \Phi _1 ^\dagger \Phi _1 ) ^2 + \lambda _2 ( \Phi _2 ^\dagger \Phi _2 ) ^2 \\
& + \lambda _3 ( \Phi _1 ^\dagger \Phi _1 ) ( \Phi _2 ^\dagger \Phi _2 ) + \lambda _4 ( \Phi _1 ^\dagger \Phi _2 ) ( \Phi _2 ^\dagger \Phi _1 ),
\end{split}
\end{equation}with the $\Phi _1$ charge freely defined. \\

Since the scalar  $\Phi_1$ plays no role, in some models such scalar is assumed not to develop a vacuum expectation value as happens in the so-called scotogenic model \cite{Farzan:2012sa,Toma:2013zsa,Hugle:2018qbw}. It is nice to see that generally considering 2HDM-$U(1)_X$ models, one can find situations where such models mimic other well-known models in the literature. The key difference between them would be the presence of a $Z^\prime$ field which is subject to interesting phenomenology \cite{Camargo:2018klg}.\\

\subsection{Right-handed neutrinos and scalar singlet absent - type II seesaw ({\bf BM 6})}
\label{sub_sec_5}

In this setup only the first term in equation (\ref{yukawa_lag_neutrino_masses}) is present, so that the matrix $ M _\nu $ degenerates to a $3 \times 3$ matrix, $M _\nu = M _L$. The neutrino masses are given simply by,
\begin{equation}
\label{mass_neutr_type_II_seesaw}
m = \sqrt{2} y  ^L v _t .
\end{equation}

In this case the potential is uniquely determined with,
\begin{widetext}
\begin{equation}
\begin{split}
\label{potential_typeII_pure}
V _H & = m _1 ^2 \Phi _1 ^\dagger \Phi _1 + m _2 ^2 \Phi _2 ^\dagger \Phi _2 + m _t ^2 \text{Tr} ( \Delta ^\dagger \Delta ) + \lambda _1 ( \Phi _1 ^\dagger \Phi _1 ) ^2 + \lambda _2 ( \Phi _2 ^\dagger \Phi _2 ) ^2 + \lambda _t [ \text{Tr} ( \Delta ^\dagger \Delta ) ] ^2 \\
& + \lambda _{tt} \text{Tr} ( \Delta ^\dagger \Delta ) ^2 + \lambda _3 ( \Phi _1 ^\dagger \Phi _1 ) ( \Phi _2 ^\dagger \Phi _2 ) + \lambda _4 ( \Phi _1 ^\dagger \Phi _2 ) ( \Phi _2 ^\dagger \Phi _1 ) + \lambda _{t1} ( \Phi _1 ^\dagger \Phi _1 ) \text{Tr} ( \Delta ^\dagger \Delta ) \\
& + \lambda _{t2} ( \Phi _2 ^\dagger \Phi _2 ) \text{Tr} ( \Delta ^\dagger \Delta ) + \lambda _{tt1} \Phi _1 ^\dagger \Delta \Delta ^\dagger \Phi _1 + \lambda _{tt2} \Phi _2 ^\dagger \Delta \Delta ^\dagger \Phi _2 + \mu _{t2} ( \Phi _2 ^T i \sigma ^2 \Delta ^\dagger \Phi _2 + h.c. ) .
\end{split}
\end{equation}
\end{widetext}where the $\Phi _1$ charge is free. The neutrino phenomenology of a pure type II seesaw has been carried out elsewhere \cite{Camargo:2019ukv}.  

\subsection{Right-handed neutrinos absent - Type II seesaw + singlet ({\bf BM 5})}
\label{sub_sec_6}

Similarly to the type II case, the neutrino masses are generated only by the scalar triplet. Therefore the expression for the neutrinos masses is the same as in Eq.\eqref{mass_neutr_type_II_seesaw}.\\

Concerning the scalars,  the charge of $\Phi _1$ and $\Phi _s$ are not fixed by Yukawa Lagrangian anymore, because right-handed neutrinos are absent. Fixing the value of $q _{Xs}$ and keeping $Q _{X1}$ free, we have the Hermitian part of the potential $V _H$ identical to the one in the Equation (\ref{potential_typeII_pure}), and three possibilities for $V _{NH}$. \\

{\bf (i)} For $q _{Xs} = Q _{X1} - Q _{X2}$ we find,
\begin{equation}
\begin{split}
V _{NH} & = \mu _s ( \Phi _1 ^\dagger \Phi _2 \Phi _s + h. c. ) + \mu _{t2} ( \Phi _2 ^T i \sigma ^2 \Delta ^\dagger \Phi _2 + h.c. ) \\
& + \kappa ' ( \Phi _1 ^T i \sigma ^2 \Delta ^\dagger \Phi _2 \Phi _s ^\dagger + h.c. ) ;
\end{split}
\end{equation}

{\bf (ii)} For $q _{Xs} = 2 ( Q _{X2} - Q _{X1} )$ we obtain,
\begin{equation}
V _{NH} = \mu _{t2} ( \Phi _2 ^T i \sigma ^2 \Delta ^\dagger \Phi _2 + h.c. ) + \kappa _1 ( \Phi _1 ^T i \sigma ^2 \Delta ^\dagger \Phi _1 \Phi _s + h.c. );
\end{equation}

{\bf (iii)} For $q _{Xs} = 0 $ we find,
\begin{equation}
V _{NH} = \mu _{t2} ( \Phi _2 ^T i \sigma ^2 \Delta ^\dagger \Phi _2 \Phi_s^{\dagger}+ h.c. ) + \kappa _2 ( \Phi _2 ^T i \sigma ^2 \Delta ^\dagger \Phi _2 \Phi _s + h.c. ) .
\end{equation}
In this last case, notice that the role of $\Phi _s$ is reduced because it contributes neither to neutrino masses (as there are no right-handed neutrinos) nor to $Z'$ one, because it is uncharged under $U(1) _X$ (see next section). Nevertheless, it does not mean that $\Phi _s$ is totally irrelevant, as it mixes with the other scalars and induces effects on the Higgs properties.\\

\section{Discussion}

One of the nice features of the 2HDM-$U(1)_X$ is the presence of a new gauge boson, a $Z^\prime$, which can be heavy or light and have different properties. These features are determined mostly by the charge assignments of the particles under $U(1) _X$ and by the scalar content of the model. For models that follow the charge assignment II (see Table \ref{table_seesaw_types_2hdm}), there are only two nontrivial particular charge assignments: one in which $d=0$ in Eq.\eqref{expres_cargas_u}, i.e. where all fermions are neutral under $U(1) _X$; and another where $d=-2/3$ that leads to fermions with $U(1) _X$ charges identical to the SM hypercharge. Different values for $d$ are in fact not distinct from the case $d = - 2/3 $, because it represents simply a rescaling on the $U(1)_X$ gauge coupling, $g _X$. Therefore, charge assignment II gives rise either to a fermiophobic or a sequential $Z'$ \cite{Camargo:2018uzw}.  We emphasize that collider bounds on such sequential $Z^\prime$ bound are rather stringent, excluding $Z^\prime$ masses below $\sim5$~TeV \cite{Aad:2019fac}, and future projection for the LHC upgrade expects to rules masses up to $10$~TeV \cite{CidVidal:2018eel}.\\ 

For the models that follow charge assignment I, the freedom in $u$ and $d$ charges in Eq. \eqref{expres_cargas_u_d} yields more possibilities, including the fermiophobic and sequential $Z'$ of the previous case, but also, a multitude of other cases, like fermiophilic $Z'$, $X = B-L$, etc \cite{Campos:2017dgc}. A detailed phenomenology of these models is outside the scope of the present work, but some of the cases of interest are discussed in \cite{Campos2017,Ko:2012hd}). \\

It is important to highlight that when there are scalar doublets, like in the model of section \ref{sub_sec_4} ({\bf BM 4}), the $Z'$ mass tends to be of the same order of the $Z$ mass or smaller, given that the $vev$ of the doublets cannot be arbitrarily large, since $v _1 ^2 + v _2 ^2 = (246 \text{\ GeV}) ^2$. In order to evade the collider bounds $g_X$ must very small because for a sufficiently light $Z^\prime$ boson, LHC loses sensitivity.\\

In the case of sections \ref{sub_sec_2} and \ref{sub_sec_5} ({\bf BM 3} and {\bf BM 6}) in which the triplet is included besides the doublets, the condition from the $W$ boson mass reads $ v ^2 + 4 v _t ^2 = (246 \text{\ GeV}) ^2$. However, the contribution of $v _t$ to the $Z'$ mass is rather restricted because of the bound from the $\rho$ parameter \cite{Camargo:2018uzw,pdg2016},
$v _t < 2 \text{\ GeV}$. Therefore, in all the cases in which there are only doublets and the triplet, the $Z'$ is necessarily light. In particular, for a $Z^\prime$ lighter than $Z$, we can generally write,

\begin{equation}
\label{light_z_prime_mass}
m _{Z '} ^2 = \frac{g _X ^2}{4 v ^2} [ ( Q _{X1} - Q _{X2} ) ^2 v _1 v _2 + q _{Xs} ^2 v _s ^2 ] ( v ^2 - 4 v _t ^2 ) .
\end{equation}
Notice that, even with the presence of the scalar singlet, $Z'$ can be light as long as $v _s$ is not so large and $g_X$ is very small. We stress that this expression for a light $Z'$ can be applied to the several specific cases treated above by setting to zero the $vev$ of the corresponding scalar that is absent. Interestingly, for sufficiently low mass, $Z'$ can behave like a dark photon \cite{Alexander:2016aln}, when $g _X$ is small and the $Z'$ interactions with fermions is dominated by the kinetic mixing term $\epsilon / 2 F ^{\mu \nu} F _{\mu \nu} '$. For $g _X$ not so small, the $Z'$ is allowed to have more general interactions with fermions. \\

As the $vev$ of the singlet is unconstrained from above, this means that $Z'$ can be made very heavy and easily evade LHC bounds that lie at the TeV scale. In this case, the contribution of $v _s$ dominates and we can approximate,

\begin{equation}
m _{Z '} = \frac{1}{2} g _X q _{Xs} v _s .
\end{equation}

Hence, as long as $v_s$ is sufficiently large, we can easily accommodate a heavy $Z^\prime$ in our model.  In summary, the models we discussed can be made consistent with existing bounds, while featuring a light or heavy $Z^\prime$.  We have not mentioned the bounds on the scalar fields masses, but they may also be circumvented by considering $v_t$ sufficiently small and $v_s$ sufficiently large. It has been shown that such bounds can be indeed evaded in the alignment limit where the mixing between the SM Higgs and the other scalars is suppressed \cite{Camargo:2019ukv}. That can be done in the models proposed here with no prejudice to neutrino masses. For this reason, we did not dwell on them.\\

\section{Conclusions}

Two Higgs Doublet Model are popular extensions beyond the SM. There are interesting alternatives to the canonical Two Higgs Doublet model such as the 2HDM-U(1) that features an additional abelian gauge symmetry. This gauge symmetry suffices to explain the absence of flavor changing neutral interactions, the presence of massive active neutrinos and dark matter. As far as neutrino masses are concerned, we proposed models that can successfully realize combinations of the type I and/or type II seesaw. We have shown that some possibilities are already excluded by data, while others remain viable, containing either relatively light or very heavy right-handed neutrinos.\\

The models we discussed here encompass other models proposed in the literature in some limiting cases, with appealing differences, such as the presence of a dark photon or heavy $Z^\prime$ gauge boson \cite{Baum:2018zhf,Iguro:2018qzf,Botella:2018gzy,Li:2018rax,Li:2018aov,Hashemi:2018gfo,Rodejohann:2019izm}.\\

We believe that such models stand as plausible alternatives to the Two Higgs Doublet Model because they are theoretically compelling for being able to address neutrino masses, dark matter and the absence of flavor changing interactions, and experimentally attractive for being subject to searches for right-handed neutrinos, dark matter, doubly charged scalars, dark photon or $Z^\prime$ fields.\\

\section*{Acknowledgements}

The authors thank Clarissa Siqueira, Antonio Santos, Daniel Camargo and Carlos Pires for discussions. FSQ thanks Diego Restrepo, Oscar Zapata and Carlos Yaguna for their hospitality at Universidad de Antioquia and UPTC Tunja U., during final stages of this work. TM thanks CAPES for the fellowship. DC thanks the support of CNPq grant 436692/2018-0. FSQ acknowledges support from CNPq grants 303817/2018-6 and 421952/2018-0, UFRN, MEC and ICTP-SAIFR FAPESP grant \#2016/01343-7. RDM acknowledges support from FAPESP grant \#2016/01343-7 and FAPESP grant \#2013/01907-0. We thank the High Performance Computing Center (NPAD) at UFRN for providing computational resources

\section{Appendix}

In this appendix we will describe in more detail the limiting cases of type I and type II seesaw dominance considering different scales for the $m_R$, $m_D$ and $m_L$ masses. We will consider the case in which we have the complete scalar sector, with the two doublets $\Phi _i$, the triplet $\Delta$ and the singlet $\Phi _s$, so that it is necessary to analyze the full mass matrix Eq.\eqref{nu_mass_matrix_simplyfied}. As we assume that the block matrix components of $M _\nu$ have equal diagonal elements, we obtain degenerate eigenvalues given by,
\begin{equation}
\label{ml}
m = \frac{1}{2} \left[ m _L + m _R - \sqrt{ 4 m _D ^2 + ( m _L - m _R ) ^2 } \right] ,
\end{equation}
and,
\begin{equation}
\label{mr}
M = \frac{1}{2} \left[ m _L + m _R + \sqrt{ 4 m _D ^2 + ( m _L - m _R ) ^2 } \right] .
\end{equation}
\\
As there are different limits, we can classify them based on the relative size of $m _L, m _R$ and $m _D$:

\begin{itemize}

\item[{\bf (i)}] The three variables are of the same order of magnitude:
\begin{equation*}
m _D \sim m _R \sim m _L .
\end{equation*}
 
\item[{\bf (ii)}] The three variables are of different orders:
\begin{equation*}
\begin{split}
& m _R \gg m _D \gg m _L , \\
& m _R \gg m _L \gg m _D , \\
& m _L \gg m _D \gg m _R , \\
& m _L \gg m _R \gg m _D , \\
& m _D \gg m _R \gg m _L , \\
& m _D \gg m _L \gg m _R .
\end{split}
\end{equation*}

\item[{\bf (iii)}] Two of them are of the same order and the third one is much larger than the others:
\begin{equation*}
\begin{split}
& m _R \gg m _D, m _L \text{\ \ and \ } m _D \sim m _L , \\
& m _L \gg m _D, m _R \text{\ \ and \ } m _D \sim m _R , \\
& m _D \gg m _R, m _L \text{\ \ and \ } m _R \sim m _L .
\end{split}
\end{equation*}

\item[{\bf (iv)}] Two of them are of the same order and the third one is much smaller than the others:
\begin{equation*}
\begin{split}
& m _L \ll m _R, m _D \text{\ \ and \ } m _D \sim m _R , \\
& m _R \ll m _L, m _D \text{\ \ and \ } m _D \sim m _L , \\
& m _D \ll m _R, m _L \text{\ \ and \ } m _R \sim m _L.
\end{split}
\end{equation*}

\end{itemize}

Instead of considering all these possibilities, we can deal with a reduced number of them, by noting that the masses $m$ and $M$ are symmetrical under the exchange of $m _R$ and $m _L$. So, we are left with:

\begin{itemize}

\item[{\bf (i)}] The three variables are of the same order of magnitude:
\begin{equation*}
m _D \sim m _R \sim m _L .
\end{equation*}

\item[{\bf (ii)}] The three variables are of different orders:
\begin{equation*}
\begin{split}
& m _R \gg m _D \gg m _L , \\
& m _R \gg m _L \gg m _D , \\
& m _D \gg m _R \gg m _L .
\end{split}
\end{equation*}

\item[{\bf (iii)}] Two of them are of the same order and the third one is much larger than the others:
\begin{equation*}
\begin{split}
& m _R \gg m _D, m _L \text{\ \ and \ } m _D \sim m _L , \\
& m _D \gg m _R, m _L \text{\ \ and \ } m _R \sim m _L .
\end{split}
\end{equation*}

\item[{\bf (iv)}] Two of them are of the same order and the third one is much smaller than the others:
\begin{equation*}
\begin{split}
& m _R \ll m _L, m _D \text{\ \ and \ } m _D \sim m _L , \\
& m _D \ll m _R, m _L \text{\ \ and \ } m _R \sim m _L.
\end{split}
\end{equation*}

\end{itemize}
The remaining cases are obtained by swapping $m _R$ and $m _L$ in the corresponding expressions. \\

For the case $m _D \sim m _R \sim m _L$, the Eq. \eqref{ml} and Eq.\eqref{mr} should be used without modification, as they are not amenable to simplifications in this regime. Now, for $ m _R \gg m _D , m _L $, we can use the approximation:
\begin{equation*}
\begin{split}
\sqrt{ 4 m _D ^2 + ( m _L - m _R )^2 } & \simeq m _R \left( 1 + \frac{m _D ^2}{m _R ^2} + \frac{m _L ^2}{ 2 m _R ^2} - \frac{m _L}{m _R} \right) \\
& = m _R - m _L + \frac{m _D ^2}{m _R} + \frac{m _L ^2}{2 m _R} .
\end{split}
\end{equation*}

Then, using Eq. \eqref{ml} and Eq.\eqref{mr}, we get, 

$\bullet$ If $ m _D \sim m _L $, then,
\begin{equation}
m \simeq m _L - \frac{m _D ^2}{2 m _R} - \frac{m _L ^2}{4 m _R} ,
\end{equation}
and,
\begin{equation}
M \simeq m _R + \frac{m _D ^2}{2 m _R} + \frac{m _L ^2}{4 m _R} .
\end{equation}
\\ \\
$\bullet$ If $ m _D \gg m _L $, then,
\begin{equation}
m \simeq m _L - \frac{m _D ^2}{2 m _R} ,
\end{equation}
and,
\begin{equation}
M \simeq m _R + \frac{m _D ^2}{2 m _R} .
\end{equation}
\\ \\
$\bullet$ If $ m _L \gg m _D $, then,
\begin{equation}
m \simeq m _L - \frac{m _L ^2}{4 m _R} ,
\end{equation}
and,
\begin{equation}
M \simeq m _R + \frac{m _L ^2}{4 m _R} .
\end{equation}
\\ \\
Now, for $ m _D \gg m _R , m _L$ we use the approximation:
\begin{equation*}
\begin{split}
\sqrt{ 4 m _D ^2 + ( m _L - m _R ) ^2 } & \simeq 2 m _D \left[ 1 + \frac{( m _L - m _R ) ^2}{8 m _D ^2} \right] \\
& = 2 m _D + \frac{( m _L - m _R ) ^2}{4 m _D} .
\end{split}
\end{equation*}

Hence:\\
$\bullet$ If $ m _R \sim m _L $, then,
\begin{equation}
m \simeq - m _D + \frac{m _L + m _R}{2} - \frac{( m _L - m _R ) ^2}{8 m _D} ,
\end{equation}
and,
\begin{equation}
M \simeq m _D + \frac{m _L + m _R}{2} + \frac{( m _L - m _R ) ^2}{8 m _D} .
\end{equation}
\\ \\
$\bullet$ If $ m _R \gg m _L $, then,
\begin{equation}
m \simeq - m _D + \frac{1}{2} m _R - \frac{m _R ^2}{8 m _D} ,
\end{equation}
and,
\begin{equation}
M \simeq m _D + \frac{1}{2} m _R + \frac{m _R ^2}{8 m _D} .
\end{equation}
\\ \\

Now, if $ m _R \ll m _L , m _D $ and $ m _L \sim m _D $:
\begin{equation*}
\begin{split}
\sqrt{ 4 m _D ^2 + ( m _L - m _R ) ^2 } \simeq \sqrt{ 4 m _D ^2 + m _L ^2 } .
\end{split}
\end{equation*}

Hence,
\begin{equation}
m \simeq \frac{1}{2} \left[ m _L - \sqrt{ 4 m _D ^2 + m _L ^2 } \right] ,
\end{equation}
and,
\begin{equation}
M \simeq \frac{1}{2} \left[ m _L + \sqrt{ 4 m _D ^2 + m _L ^2 } \right] ,
\end{equation}
\\ \\
Finally, for $ m _D \ll m _R , m _L $ and $ m _R \sim m _L $:
\begin{equation*}
\begin{split}
\sqrt{ 4 m _D ^2 + ( m _L - m _R ) ^2 } & = - ( m _L - m _R ) \left( 1 + \frac{2 m _D ^2}{( m _L - m _R ) ^2} \right) \\
& = m _R - m _L - \frac{2 m _D ^2}{( m _L - m _R )} .
\end{split}
\end{equation*}

Hence,
\begin{equation}
m \simeq m _L - \frac{m _D ^2}{( m _R - m _L )} ,
\end{equation}
and,
\begin{equation}
M \simeq m _R + \frac{m _D ^2}{( m _R - m _L )} .
\end{equation}
\\

\bibliography{referencias}

\end{document}